\documentstyle[aps,prl,multicol,psfig]{revtex}
\tighten
\begin{document}
%\draft
\title{Relationship between incommensurability and superconductivity in
Peierls distorted charge-density-wave systems}

\author{G. Seibold and S. Varlamov}
\address{Institut f\"ur Physik, BTU Cottbus, PBox 101344, 
         03013 Cottbus, Germany}
\date{\today}
\maketitle

\begin{abstract}
We study the pairing potential induced by fluctuations around a 
charge-density wave (CDW) with scattering vector ${\bf Q}$ 
by means of the Fr\"ohlich transformation. 
For general commensurability
M, defined as $|{\bf k}+M{\bf Q}\rangle=|{\bf k}\rangle$, we find 
that the intraband
pair scattering within the M subbands scales with M  whereas the interband 
pair scattering 
becomes suppressed with increasing CDW order parameter.
As a consequence superconductivity is suppressed when the Fermi energy
is located between the subbands as it is usually the case for
nesting induced CDW's, but due to the vertex renormalization it 
can be substantially enhanced when the
chemical potential is shifted sufficiently inside one of the subbands. 
The model can help to understand the experimentally observed 
dependence of the superconducting transition temperature from the 
stripe phase incommensurability in the lanthanum cuprates.
\end{abstract}

\vspace*{0.2cm}


\begin{multicols}{2}
\section{Introduction}
It is now well established that some of the cuprate high-T$_c$
superconductors show a strong depression of $T_c$ when the
doping concentration coincides with an integer fraction.
This feature was first detected in La$_{2-x}$Ba$_{x}$CuO$_4$ 
\cite{MOODENBAUGH} where the superconducting transition temperature
displays a clear minimum near compositions $x \approx 0.12 \approx 1/8$.  
Subsequent measurements on La$_{2-x}$Sr$_{x}$CuO$_{4}$ (LSCO) also
resulted in a detectable dip of $T_c$ around the same doping
concentrations \cite{TAKA1} although the reduction is less
pronounced in this system than in the Ba-doped samples..
Further insight into this anomaly was provided by inelastic neutron
scattering measurements on La$_{2-x}$Sr$_{x}$CuO$_{4}$ (LSCO) 
\cite{CHEONG,MASON,THURST} which revealed the existence of 
incommensurate spin correlations in this material.   
More recently it was found that these incommensurate spin fluctuations
are pinned in Nd doped LSCO oxide compounds 
\cite{TRAN,TRAN2} 
leading to the appearance of magnetic {\it and} charge-order superlattice
peaks in the neutron diffraction pattern.
These measurements therefore provide strong evidence for the occurence
of spin- and charge-stripe order in the lanthanum compounds which becomes
static when the lattice undergoes a transition to a low temperature
tetragonal (LTT) phase as it is the case for the Ba- and Nd doped systems.
However, there is now growing evidence that this structural phase
transition is not a necessary condition for the occurence of
static stripe order which recently has also been detected in
LSCO \cite{KIMURA} and oxygen-doped La$_2$CuO$_{4+y}$ \cite{LEE}.
It should be noted that in case of the LSCO compound, ultrasonic
measurements carried out for doping $x=0.12$ show a mode softening 
below $\sim 45$ K \cite{SUZUKI}, which can be understood as a precursor 
of the LTT structural phase transition.
This in turn can act as the pinning mechanism for the fluctuating stripe
order and therefore explain the small dip in $T_c$ around $x \approx 1/8$ 
for the LSCO material.

With regard to the origin of incommensurable scattering
it has been concluded that the stripe
ordering is driven by the charges \cite{TRAN,NIEMOELL}
since the magnetic peaks appear at lower temperatures
than the charge-order peaks,
consistent with an Landau free-energy
analysis of coupled charge- and spin-density order parameters \cite{ZACHAR}.
For completeness we want to mention that incommensurate 
spin correlations have also been detected in the YBCO 
compounds \cite{MOOK} and the existence of charge modulation
in the CuO$_2$ planes of Bi2212 was proposed early on in Ref. \cite{BIANC}.

It is therefore naturally to ask wether there may exist a connection
between the occurence of stripe modulations and the superconducting
pair formation. A commonly held opinion is that 
stripe correlations naturally lead to a depression
of T$_c$ (as observed in Nd-doped LSCO) due to the (nesting induced) 
reduction of the density of states at the Fermi level and the
localization of the charge carriers through the pinning potential
of a possible structural phase transition.
However, most interesting results have been obtained in the LSCO system where
both the incommensurate spin fluctuations and superconductivity have been
studied as a function of doping \cite{YAMADA}.
It turned out that the incommensurability of the magnetic fluctuations
(defined as the deviation of the magnetic peak from its
antiferromagnetic position) scales linearly with the superconducting
transition temperature T$_c$ up to optimal doping, i.e. both $\delta$
and $T_c$ appear at the same critical concentration x$\approx 0.05$ and follow
the same doping dependence. This result strongly suggests that incommensurate
stripe ordering and superconductivity may arise from the same instability
as it is proposed in the framework of a Quantum Critical Point (QCP) scenario
\cite{CAST}. Within this model the singular scattering induced by the
critical charge fluctuations would be responsible for both the anomalous
normal-state properties and the large superconducting critical temperatures.
Locating the QCP near optimal doping the phase diagram of the cuprates is 
partitioned in a (nearly) ordered, a quantum critical, and a quantum
disordered region naturally corresponding to the under-, optimally, and
over-doped regions of the phase diagram of the cuprates. 

Presupposing the existence of charge stripe correlations we want
to investigate in the present paper how the incommensurability
of these modulations may affect the pairing potential. For
convenience we restrict ourselves to the static case using a Fr\"ohlich
type transformation to study the
problem how the modification of the electron states
due to the formation of a charge-density wave (CDW) influences the 
superconducting order parameter for given (in)commensurability.
It should be mentioned that a similar model has already been considered
in Ref. \cite{BALSEIRO} where the effective electron-electron interaction  
mediated by phonons has been factorized due to CDW and superconducting
order parameters. However, this procedure is not consistent
with the formation of a CDW since it leads to an unnatural energy
cutoff for the particle-hole excitations whereas a {\it static}
CDW involves particle-hole scattering processes of all available
energies. The correct procedure which we will follow below is therefore
first to diagonalize the CDW mean-field part of the hamiltonian and
after that to calculate the effective interaction between the new
quasiparticles arising from fluctuations around the frozen-in CDW.
It turns out that this interaction strongly depends on the 
(in)commensurability M of the CDW with the underlying lattice 
which opens the possibility of increasing T$_c$ by doping the
CDW while fixing its periodicity. We want to stress that
this mechanism is {\it not} due
to the appearance of $1/\sqrt{\omega}$-singularities in the density of
states (DOS) but has its origin in the modification of the electron-phonon
vertex function. 

In Sec. II we introduce our model hamiltonian and in Sec. III we will 
derive the effective quasiparticle interaction in a two-dimensional 
CDW system of arbitrary
commensurability using a Fr\"ohlich transformation approach. 
Explicit results are presented in Sec. III for the simplest case
of a CDW with modulation ${\bf Q}=(\pi,\pi)$. 
Despite its low commensurability
it turns out that T$_c$ is  considerably enhanced due to the vertex
renormalization when the chemical potential is located sufficiently inside one
of the two subbands.
We finally summarize our conclusions in Sec. IV.

\section{Model Hamiltonian}
Our investigations are based on the following lattice model which describes
the coupling of electrons to a dispersionless phonon mode
\begin{eqnarray}\label{e1}
H&=&\sum_{k,\sigma}\epsilon_{k} c_{k,\sigma}^{\dagger}c_{k,\sigma}
 + \omega_0 \sum_q b_q^{\dagger}b_q \nonumber \\
&+&\frac{g}{\sqrt{N}}\sum_{kq,\sigma}c_{k+q,\sigma}^{\dagger}c_{k,\sigma}
(b_q+b_{-q}^{\dagger})
\end{eqnarray}
where $c_{k,\sigma}^{(\dagger)}$ destroys (creates) an electron in the
state k and energy $\epsilon_k$. The operatos $b_{q}^{(\dagger)}$ 
destroy (create) a phonon with momentum {\it q}.
For convenience we take the electron-phonon
matrix element g and the phonon frequency $\omega_0$ to be constant.  

Within this model the transition to a static CDW 
is signaled by a diverging polarizability for a certain wave vector
Q and at some critical doping concentration.
The phonon part of the CDW mean-field wave-function is usually
taken to be a coherent state which means that for the selected
Q-values  the phonon operators are replaced by a complex number.

The coherent state contribution is extracted from the
hamiltonian Eq. (\ref{e1}) by means of a displacement transformation
$b_q=a_q-\alpha_q$ where the operators $a_q$ now correspond to the
phonon fluctuations around the coherent state $|\alpha_q\rangle$. 
The CDW order parameter is defined through $g \alpha_q / \sqrt{N}=\Delta_Q
\delta_{q,Q}$. We have neglected the contribution of
higher harmonics to the CDW profile which are not essential
for our present considerations.

We thus obtain the following hamiltonian
\begin{equation}\label{HH}
H=H_{CDW}+H_{Fl}+H_{P}
\end{equation}
where
\begin{eqnarray}\label{HMF}
H_{CDW}&=&\sum_{k,\sigma}\epsilon_{k} c_{k,\sigma}^{\dagger}c_{k,\sigma}
+N \frac{\omega_0}{g^2} |\Delta_Q|^2
\nonumber \\
&-&\sum_{k,\sigma} \left(\Delta_Q c_{k+Q,\sigma}^{\dagger}c_{k,\sigma}
+\Delta_Q^{*}c_{k-Q,\sigma}^{\dagger}c_{k,\sigma}\right)
\end{eqnarray}  
decribes the static CDW scattering with wave vector Q of the electrons.
The essential term which contains the coupling of the phonon fluctuations
to the electronic degrees of freedom corresponds to
\begin{equation}\label{HCP}
H_{FL}= \frac{g}{\sqrt{N}}\sum_{kq,\sigma}c_{k+q,\sigma}^{\dagger}c_{k,\sigma}
(a_q+a_{-q}^{\dagger})
\end{equation}
and 
\begin{equation}\label{hp}
H_{P}=\omega_0 \sum_q a_q^{\dagger}a_q
\end{equation}
is the energy of the phonon fluctuations.

\section{Effective interaction for general commensurability}
For a given k-state in the Brillouin zone we define the 
commensurability M of the CDW as the number of scattering
events in the same direction needed to return to the equivalent state, i.e.
$|{\bf k}+M{\bf Q}\rangle=|{\bf k}\rangle$. It is clear that the larger
M the higher is the 'incommensurability' of the CDW with the underlying
lattice.
In order to keep the comprehensibility of our presentation we consider
in the following a two-dimensional system where the CDW scattering 
vector ${\bf Q}=(Q_x,0)$ is
along the x-direction. Within the derivation presented below other 
orientations would only change the k-state labeling. 
The reduced Brillouin zone (BZ) then is defined by 
$-\pi/M < k_x \le \pi/M$
and $-\pi/a < k_y \le \pi/a$ and the electron operators in the 
reduced zone read as 
\begin{equation}
c_n(k,\sigma)\equiv c_{k_x-{\rm sgn}(k_x)(n-1)Q_x, k_y, \sigma} 
\end{equation}
where the same definition holds also for the phonon operators.
The signum function has been introduced in order to obtain symmetric
subbands in the reduced zone with respect to $k_{x}$.
Performing a unitary transformation $c_n(k,\sigma)=\sum_{m=1}^M A_{nm}
f_m(k,\sigma)$ the mean-field part of the hamiltonian Eq. (\ref{HMF})
can be diagonalized 
\begin{equation}
H_{CDW}=\sum_{k,\sigma}\sum_{n}E_n(k)f_n^{\dagger}(k,\sigma)f_n(k,\sigma)
\end{equation}
where $E_n(k)$ denotes the dispersion of the n-th subband in the reduced BZ.
It can be further shown that the condition $E_n(k_x,k_y)=E_n(-k_x,k_y)$
requires for the transformation matrices the property 
$A_{nm}(k_x,k_y)=A^{*}_{nm}(-k_x,k_y)=A^{-1}_{mn}(-k_x,k_y)$.

The coupling term Eq. (\ref{HCP}) rewritten in terms of the
new quasiparticle operators reads as
\begin{eqnarray}\label{HTC}
H_{FL}&=&\frac{g}{\sqrt{N}} \sum_{kq,\sigma}\sum_{stm}
\Gamma_{stm}(k,q)f_s^{\dagger}(\widetilde{k+q},\sigma)f_{t}(k,\sigma)
\nonumber \\ &\times& \left(a_m(q)+a^{\dagger}_m(-q)\right)
\end{eqnarray}
where we have introduced the vertex function
\begin{equation}\label{VF}
\Gamma_{stm}(k,q)= \sum_n A^{*}_{{\cal N},s}
(\widetilde{k+q})A_{n,t}(k)
\end{equation}
and the index ${\cal N}={\cal N}(n,m,k,q)$ is defined below.
Unfortunately at this point the notation is getting complex both 
due to the addition of momenta in the signum function and the scattering
of electrons between neighboring reduced zones.
For this reason the following definitions have been made in Eqs. (\ref{HTC},
\ref{VF})
\begin{eqnarray}
{\rm For}&\quad& {\rm sgn}(k_x)={\rm sgn}(q_x) 
\quad{\rm and}\quad |k_x+q_x|<\pi/M : \nonumber \\
\widetilde{k+q}&\equiv& k+q ;\quad {\cal N}=n+m-1 \label{DF1} \\
{\rm For}&\quad&{\rm sgn}(k_x)\neq{\rm sgn}(q_x) 
\quad{\rm and}\quad |k_x+q_x|<\pi/M : \nonumber \\ 
\widetilde{k+q}&\equiv& k+q ;\quad {\cal N}={\rm sgn}(|k_x|-|q_x|)(n-m)+1 
 \label{DF2}\\
{\rm For}&\quad& {\rm sgn}(k_x)={\rm sgn}(q_x) 
\quad{\rm and}\quad |k_x+q_x|>\pi/M : \nonumber \\
\widetilde{k+q}&\equiv& k+q-{\rm sgn}(k_x+q_x)Q_x ;\quad {\cal N}=n+m-2 
\label{DF3}
\end{eqnarray}

The coupling term Eq. (\ref{HTC}) can now be eliminated in first order
by means of a standard Fr\"ohlich transformation (see e.g. \cite{WAGNER}), 
i.e. we expand the hamiltonian Eq. (\ref{HH}) in a commutator series
up to second order in the vertex $\Gamma$
\begin{equation}
\tilde{H}={\rm e}^{-S}H{\rm e}^{S}=H+[H,S]+\frac{1}{2}[H,S],S] + \cdots
\end{equation}
with S being specified by the condition:
\begin{equation}\label{COND}
[H_{CDW}+H_{P},S]=-H_{Fl}
\end{equation}
Representing S in the form
\begin{eqnarray}
S&=&\frac{1}{\sqrt{N}} \sum_{kq,\sigma}\sum_{uvw}f_u^{\dagger}
(\widetilde{k+q},\sigma)f_v(k,\sigma)\nonumber \\
&\times& \lbrack P_{uvw}^{-}(k,q)a_w(q)+P^{+}_{uvw}(k,q)a_w^{\dagger}(-q)
\rbrack
\end{eqnarray}
the requirement Eq. (\ref{COND}) is satisfied by 
\begin{equation}
P_{uvw}^{\pm}(k,q)=- g \frac{\Gamma_{uvw}(k,q)}
{E_u(\widetilde{k+q})-E_v(k) \pm \omega_0}
\end{equation}

We thus finally obtain a transformed hamiltonian which, among terms
describing the renormalization of phonon and quasiparticle energies, 
contains an effective quasiparticle interaction.
The relevant contribution for superconductivity consists of those
terms which describe the scattering of a zero momentum pair between
the subbands of the CDW system and reads as
\begin{eqnarray}\label{VEFF}
V_{SC}&=&\frac{1}{N}\sum_{kq,\sigma}\sum_{stm}|\Gamma_{stm}(k,q)|^2
\frac{g^2 \omega_0}{[E_s(\widetilde{k+q})-E_t(k)]^2-\omega_0^2} \nonumber \\
&\times& f_{s}^{\dagger}(\widetilde{k+q},\sigma)
f_{s}^{\dagger}(\widetilde{-k-q},-\sigma)f_t(-k,-\sigma)f_t(k,\sigma)
\end{eqnarray}

Obviously this expression reduces to the well-known phonon mediated
electron-electron interaction in the limit 
$\Delta_Q \rightarrow 0$ when the transformation
matrices are given by $A_{nm}(k)=\delta_{n,m}$ and the vertex contribution
becomes independent of the band indices 
$\sum_{m}|\Gamma_{stm}(k,q)|^2 \rightarrow 1$.
However, in case of a well developped CDW the transformation 
takes the form 
\begin{eqnarray}
A_{np}(k)&\approx& \frac{1}{\sqrt{M}}
\exp{(i \,n\,p\,{\rm sgn}(k_x))}\\
 p&=&-\pi/M \dots \pi/M \nonumber
\end{eqnarray}
which for the
vertex function leads to $\Gamma_{stm}(k,q)=\exp{(i\phi(m))}\delta_{s,t}$
i.e. the phonon branch-index m only enters through a phase $\phi(m)$
depending on the three ranges defined in 
Eqs. (\ref{DF1}-\ref{DF3}).
As a consequence the vertex contribution
\begin{equation}
\sum_m |\Gamma_{stm}(k,q)|^2 = M \delta_{s,t}
\end{equation}
now restricts the pair-pair scattering to the individual subbands and
the effective interaction becomes proportional to the
commensurability M of the CDW.
Since the number of states within each subband decreases with
the commensurability as $2N/M$ it is obvious that the vertex
renormalization can only lead to an effective enhancement of the 
quasiparticle interaction as long as the cutoff range of its attractive
contribution lies within one of the subbands. For the present two-dimensional
model with a phonon mediated interaction this means that for
$M<B/\omega_D$ the CDW formation can lead to a substantial
enhancement of the superconducting order parameter where B is the
bandwidth and $\omega_D$ denotes the Debye frequency as a characteristic
energy scale for the cutoff. 
Of course the considered limit of a large CDW order parameter is rather 
academic since then the quasiparticles become localized and superconductivity
will be suppressed.
However, the essential point is that with the onset of CDW scattering
the intraband pairing is amplified with a simultaneous decrease of the 
interband pairing amplitude. Both processes do {\it not} compensate 
since most of the interband contribution to the interaction 
falls out of the range of attraction.

\vspace*{1cm}

\begin{figure}
\hspace{0cm}{{\psfig{figure=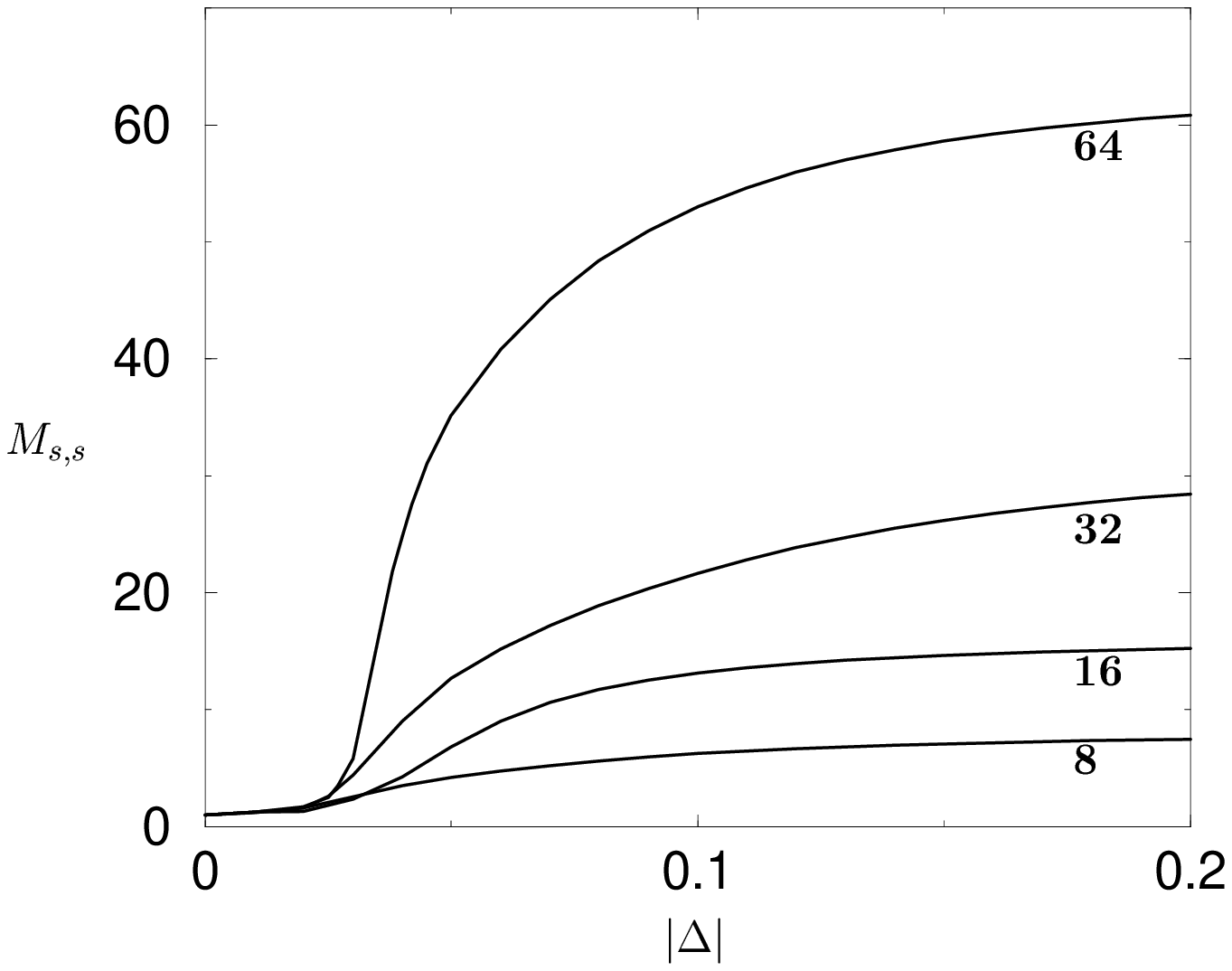,width=7cm}}}

%\vspace*{0.2cm}

{\small FIG. 1. Intraband vertex contribution to the effective interaction
$M_{s,s}=\sum_m |\Gamma_{ssm}(k,q)|^2$ for ${\bf Q}=(3\pi/4,0)$, $(5\pi/8,0)$,
$(9\pi/16,0)$, and $(19\pi/32)$ corresponding to the commensurabilities
$M=8, 16, 32$, and 64 as indicated in the figure. ${\bf k}=(\pi/80,\pi)$,
${\bf q}=(-\pi/40,0)$}
\end{figure}

In the remainder of this section we consider as an example the CDW induced
vertex renormalization for holes with energy dispersion
\begin{eqnarray*}
\epsilon_k&=&2t_1(\cos k_x + \cos k_y) + 4t_2\cos k_x \cos k_y\\
          &+& 2 t_3(\cos 2k_x + \cos 2k_y)
\end{eqnarray*}
where we use the hopping integrals $t_1=45$meV, $t_2=5$meV, $t_3=16$meV
\cite{EREMIN}.
For small and moderate hole concentrations this dispersion gives rise
to an open (i.e. centered around $(\pi,\pi)$) Fermi surface as observed 
in various photoemission experiments in the high-T$_c$ cuprates (see e.g.
\cite{SHEN} and references therein).
Moreover one finds \cite{EREMIN} that in the 'underdoped' regime this
model leads to a polarizability which is peaked along the $k_x$- and $k_y$-
axis of the Brillouin zone repectively. 
As a consequence the system becomes unstable
towards a vertically oriented CDW when the electron-phonon coupling constant
exceeds some critical value. In the following we consider a scattering vector
$Q \approx (0.6 \pi,0)$ which in our model is realized for hole concentrations
$\delta \approx 0.25$. 

Fig. 1 displays the intraband vertex contribution to the effective interaction
$M_{s,s}(k,q) =\sum_m |\Gamma_{ssm}(k,q)|^2$ for different commensurabilities,
i.e. the scattering vector  ${\bf Q} \approx (0.6 \pi,0)$ is approximated by
increasing fractions $(n/m,0)$ as indicated in the figure capture. 
For each case we have selected a subband which covers
the Fermi level and the phase of the CDW order parameter is fixed to
$\phi=\pi/3$. The general behavior of $M_{s,s}(k,q)$ is consistent with the
limiting values discussed above. Starting from $M_{s,s}(k,q)=1$ 
one observes a strong increase for some critical value of
the CDW order parameter $|\Delta|_{crit}$ and a saturation to the maximum
commensurability in the limit of large $|\Delta|$.
The value of $|\Delta|_{crit}$ strongly depends on the selected band and
especially on the values of ${\bf k}$ and ${\bf q}$. When these states
are choosen such that their scattered states (i.e. ${\bf k}+{\bf Q}$, 
${\bf k}+2{\bf Q}$ ...
and ${\bf q}+{\bf Q}$, ${\bf q}+2{\bf Q}$ ...) do not differ much in energy, 
the onset of the increase of $M_{s,s}(k,q)$ occurs for smaller $|\Delta|$ as
in case of large energy differences between the scattered states.

\vspace*{0.5cm}

\begin{figure}
\hspace{0.3cm}{{\psfig{figure=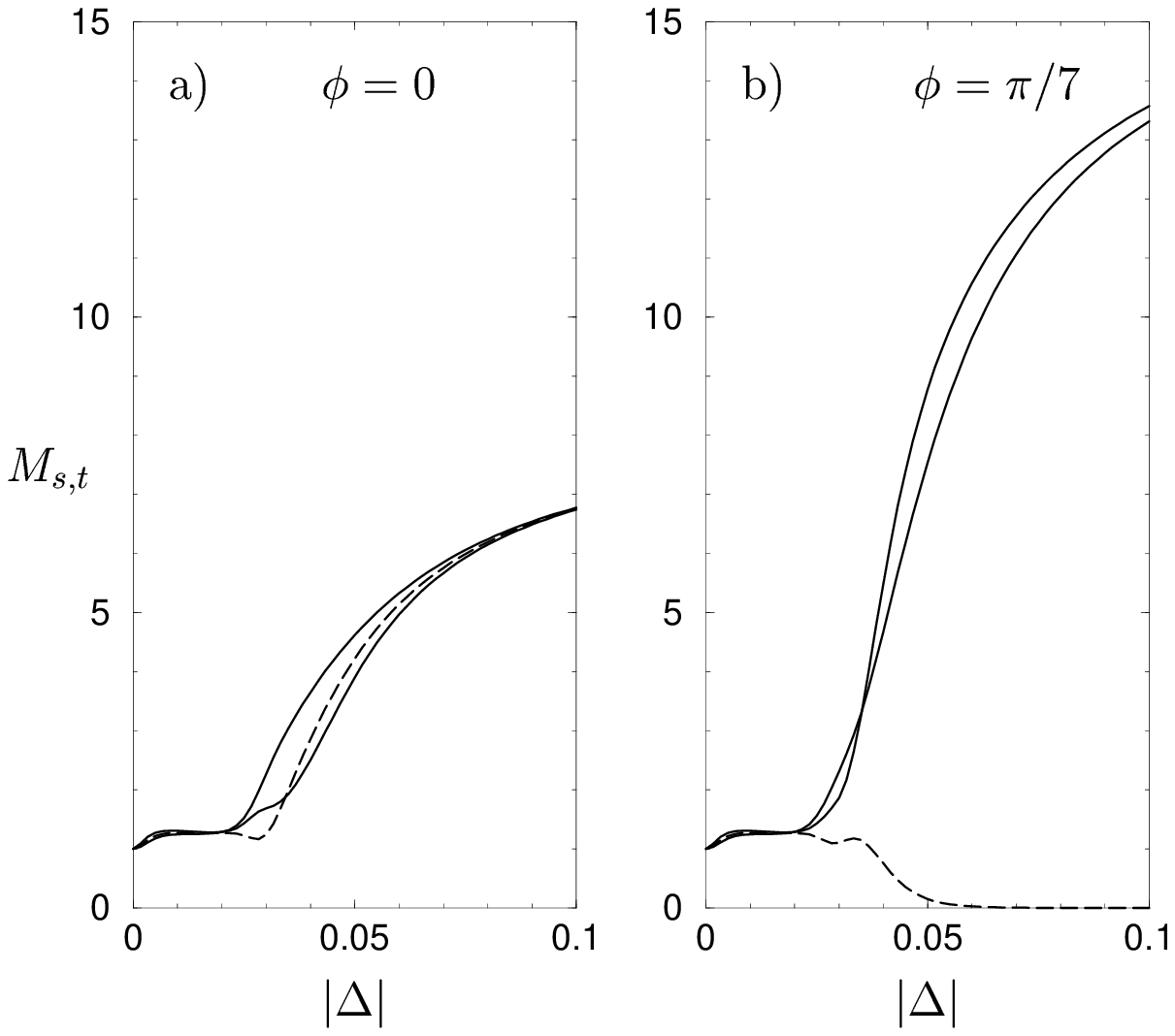,width=7cm}}}

\vspace*{0.2cm}

{\small FIG. 2. Vertex contribution to the effective interaction
$M_{s,t}=\sum_m |\Gamma_{stm}(k,q)|^2$ for two energetically neighbored
subbands of a CDW
with scattering vector ${\bf Q}=(5\pi/8,0)$ (commensurability M=16). 
The solid lines correspond to the intraband scattering ($M_{1,1}$ and
$M_{2,2}$) whereas the dashed line refers to the scattering between the
subbands ($M_{1,2}$). a) and b) are plots for two different phases
$\phi=0,\pi/7$ of the CDW order parameter $\Delta=|\Delta|\exp(i\phi)$.}
\end{figure}

Another point not discussed so far is the influence of the phase
of the CDW order parameter. In Fig. 2 we show  the vertex contribution
$M_{s,t}$ for two energetically neighbored subbands of a CDW
with scattering vector $Q=(5\pi/8,0)$ (commensurability M=16) and two
different phases $\phi=0,\pi/7$. In case of $\phi=0$  
not only the intraband scattering is enhanced with increasing $\Delta$ 
but also the scattering between the bands due to their large
degeneracy. As a result the curves for the intraband contribution approach
half the value of the commensurability only in the limit of large
$|\Delta|$, in favor of a similar behavior of the interband renormalization.  
As shown in Fig. 2b this interband scattering can be suppressed by choosing
a different phase which essentially restores the 'orthogonality'
between the quasi-degenerate states. 
However, for the quasiparticle pairing the phase of the CDW order parameter
is not relevant since it essentially redistributes the pair-scattering
between different channels each having the same pairing amplitude. 
 
\section{CDW with commensurability M=2}
We will now discuss the consequences of the renormalized vertex to
superconductivity in more detail studying
the simplest available CDW system in two dimensions with scattering
vector $Q=(\pi,\pi)$ and electron dispersion 
$\epsilon_k=-\epsilon_{k+Q}=-B/2(\cos k_x + \cos k_y)$,
corresponding to nearest neighbor hopping of particles on a square lattice.
The hamiltonian Eq. (\ref{HMF}) then can be easily diagonalized 
and one obtains
\begin{eqnarray}
H_{\rm CDW}&=&\sum_{k,\sigma}E_k\lbrack f_{1}^{\dagger}(k,\sigma)f_{1}(k,\sigma)
-f_{2}^{\dagger}(k,\sigma)f_{2}(k,\sigma)\rbrack \nonumber \\
&+& \frac{N}{4} 
\frac{\omega_0}{g^2} \Delta_{\rm CDW}^2
\end{eqnarray}
where the dispersion of the CDW system is given by 
$E_k=\sqrt{\Delta_{\rm CDW}^2+\epsilon_k^2}$.

Following the procedure described above we then derive
the pairing hamiltonian
\begin{eqnarray}\label{HEFF}
H_{\rm el-el}&=&\frac{1}{N}\sum_{kq,\sigma} \sum_{i,j}^2 \Gamma_{ij}^2(kq) 
\frac{g^2 \omega_0}{[E(k+q)-E(k)]^2-\omega_0^2}  \nonumber \\
&\times&
f_{i}^{\dagger}(-k-q,-\sigma)f_{i}^{\dagger}(k+q,\sigma)
f_{j}(k,\sigma)f_{j}(-k,-\sigma)
\end{eqnarray}
where the vertex contribution reads
\begin{equation}
\Gamma_{ij}^2(kq)=\left( 1\pm\frac{\Delta_{\rm CDW}^2}{E_k E_{k+q}}\right)
\end{equation}
and the $+$($-$) sign corresponds to $i=j$($i \neq j$).
The pair scattering between valence and conduction band thus is decreased
with increasing CDW order parameter in favor of an enhancement of
the intraband pairing amplitude. 

We have calculated selfconsistently the CDW- and SC order parameters 
within a BCS decoupling scheme
and furthermore have used a BCS-type approximation for the interaction 
in Eq. (\ref{HEFF}), i.e.
replacing $g^2 \omega_0/{[E(k+q)-E(k)]^2-\omega_0^2}$ by a constant
potential $-V$ for k-states out to a cutoff energy $\omega_D$.
The symmetry of the SC gap is then completely determined by the
symmetry of $\Gamma_{ij}^2(kq)$ and takes the form
\begin{equation}\label{GAP}
\Delta_{SC}(k)=\Delta_0 + \alpha \frac{\Delta_{\rm CDW}^2}{E_k}.
\end{equation}
\begin{figure}
\hspace{0.6cm}{{\psfig{figure=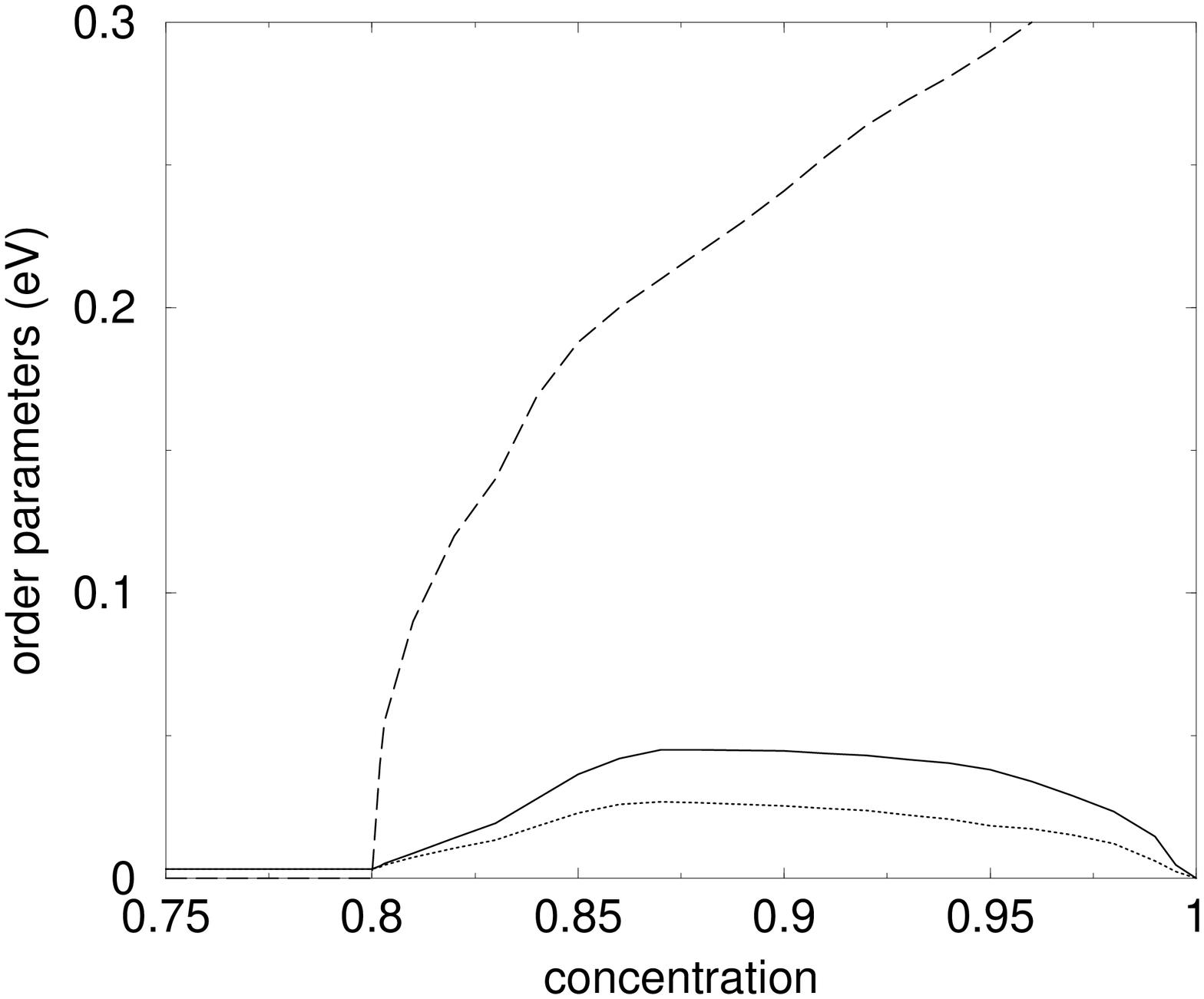,angle=-90,width=7cm}}}

%\vspace*{0.2cm}

{\small FIG. 3. CDW- and SC order parameters as a function of doping
for a $Q=(\pi,\pi)$ CDW system. Dashed line: CDW gap ($\Delta_{\rm CDW}$), 
Dotted line: constant part of the SC gap ($\Delta_0$), 
Solid: Maximum value of the total SC gap ($\Delta_{SC}^{\rm max}(k)$).
For convenience the calculation has been performed for a constant DOS
$\rho(\omega)=1/(2B)\Theta(B^2-\omega^2)$. Parameters: B=0.8eV, V=0.25eV,
$\omega_D=$0.04eV.} 
\end{figure}
Fig. 3 displays the results as a function of doping (concentration
$n=1$ corresponds to half-filling) for a selected parameter set. 
The onset of CDW scattering leads to a $1/\sqrt{\omega}$ contribution
to the DOS around the CDW gap-edges. As a consequence also for concentrations
$n < 1$ the DOS at the Fermi energy is increased leading to an enhancement
of the constant part of the SC gap.
Besides this well known effect it can be seen that the vertex renormalization
induces a considerable increase of the SC order parameter due to
the amplification of the intraband pairing.
Additionally superconductivity is suppressed when the concentration
leads to perfect nesting of the Fermi surface (i.e. $n=1$).
In the small coupling limit one can obtain an explicit expression for
the vertex contribution to the SC gap under the assumption
that the chemical potential $\mu$ is sufficiently inside the lower subband 
($\mu < -\Delta_{\rm CDW}-\omega_D$). 
We find 
\begin{equation}
\alpha = -\frac{V}{4B |\mu|}\Delta_0 \ln\lbrack \frac{\Delta_0^2}{4}
\left(1/\omega_D^2+1/\mu^2\right)\rbrack
\end{equation}
where B denotes the electronic bandwidth.
Note that $\alpha>0$ since the expression has been obtained under
the assumption $\Delta_0 < \omega_D,|\mu|$.

\section{Conclusion}
To summarize,
we have shown that the vertex renormalization can lead to a substantial
enhancement of the superconducting order parameter in a Peierls distorted
CDW system when the chemical potential is located sufficiently inside
one of the subbands. Since for a nesting induced instability the 
Fermi level always falls in the DOS minimum between two subbands it
is clear that this mechanism can only work when the CDW modulation
is fixed by some external (e.g. lattice-) potential while the carrier
concentration is varied. This in turn, however, is not in favor of 
superconductivity due to the localization of the charge carriers. 
The more promising alternative would be a nesting
independent CDW instablity as realized for example in a frustrated
phase separation scenario \cite{CAST}. In this case the CDW formation
occurs due to the interplay between long-range Coulomb forces and
a phase separation instability and is not related to any Fermi wave
vector $k_F$. 

Our considerations provide a possible explanation for the 
linear relationship between $T_c$ and incommensurability 
as observed in the underdoped lanthanum cuprates \cite{YAMADA}.
In these systems the incommensurability M is related to the
deviation of the magnetic scattering vector 
${\bf Q}=(\pi/M,\pi)$ from $(\pi,\pi)$, i.e.
upon doping towards $T_c^{max}$  one observes an increase of the 
incommensurability M. According to our findings this in turn induces 
an enhancement of the effective interaction $V_{eff} \sim M$   
which causes T$_c$ to scale with the incommensurability M. 
Of course we have used a very simplified model  
restricting ourselves to static CDW formation and neglecting
spin degrees of freedom. However the main outcome of our investigations,
namely the scaling of the effective interaction with the commensurability, 
should hold also in case of dynamic CDW scattering.
Indeed there is experimental evidence that in the
La$_{2-x}$Sr$_{x}$CuO$_4$ system,
where no pinning is provided by the LTO-LTT transition, stripes fluctuate
very slowly \cite{HUNT} allowing for an order parameter description 
in principle also in this case.

%We finally like to point out that the mechanism presented here may provide an
%explanation for  the difference in T$_c$
%between the barium bismuthates and the cuprate superconductors due
%to the different CDW instabilities realized in these materials. Whereas in
%the BaBiO$_3$ system the charge-ordered state can be viewed as a 
%(nesting induced) breathing mode distortion of commensurability 
%$M=2$, large values of
%the commensurability (possibly due to a frustrated phase separation mechanism)
%have been observed in the cuprates which according to our present analysis
%can lead to significantly enhanced transition temperatures.

\acknowledgments
We would like to thank A. Bill for helpful comments and a critical reading
of the manuscript. G.S. greatfully acknowledges stimulating discussions with 
M. Grilli, C. Castellani and C. Di Castro 
at the early stage of this work.

\end{multicols}

\end{document}